\documentclass{ws-ijmpa}

\usepackage{latexsym}
\usepackage{graphicx}

\begin{document}
   
\markboth{L.M. Diaz-Rivera \& L.O. Pimentel}
{Conditions for acceleration and deceleration in Sacalar-Tensor theories}

\catchline{}{}{}

\title{ REQUIRED CONDITIONS FOR LATE TIME ACCELERATION AND EARLY TIME DECELERATION IN GENERALIZED SCALAR-TENSOR THEORIES}

\author{ L. M. DIAZ-RIVERA\footnote{e-mail:luz@phys.ufl.edu}} 

\address{Physics Department, University of Florida,  PO Box 118440, Gainesville, FL, 32611-8440}

\author{LUIS O. PIMENTEL}

\address{Physics Department, Universidad Aut{\'o}noma Metropolitana Iztapalapa P.O. Box 55-534, 09340, M{\'e}xico D. F., M{\'e}xico}

\maketitle

\pub{Received (Day Month Year)}{Revised (Day Month Year)}

\begin{abstract}

We consider a generalized scalar-tensor theory, where we let the coupling function $\omega(\phi)$ and the effective cosmological constants $\Lambda(\phi)$ undetermined. We obtain general expressions for $\omega(\phi)$ and $\Lambda(\phi)$ in terms of the scalar field and the scale factor, and show that $\omega(\phi)$ depends on the scalar field and the scale factor in a complicated way. In order to study the conditions for an accelerated expansion at the present time and a decelerated expansion in the past, we assume a power law evolution for the scalar field and the scale factor. We analyse the required conditions that allow our model to satisfy the weak field limits on $\omega(\phi)$, and at the same time, to obtain the correct values of cosmological parameters, as the energy density $\Omega_{m0}$ and cosmological constant $\Lambda(t_0)$. We also study the conditions for a decelerated expansion at an early time dominated by radiation. We find values for $\omega(\phi)$ and $\Lambda(\phi)$ consistent with the expectations of a theory where the cosmological constant decreases with the time and the coupling function increases until the values accepted today.


\end{abstract}

\section{Introduction}

There are two recent observational results that have had a profound implication in our current understanding of the universe, and consequently both results have triggered an intense activity in cosmology; namely the confirmation that our universe is flat, and that it expands in an accelerated way. On one hand, the observed location of the first acoustic peak of the temperature fluctuations on the Cosmic Microwave Background (CMB) corroborated by the data obtained by BOOMERANG and MAXIMA \cite{bernadis}, favors a spatially flat universe and indicates that the universe is dominated by an unidentified ``dark energy''. On the other hand, direct evidences based upon measurements of luminosity-redshift relations for type Ia supernovae, carried out independently by two groups \cite{perlmutter}, indicate that the universe is presently expanding in an accelerated way, and suggest that the unidentified dark energy has a negative pressure. This last characteristic of the dark energy points to the vacuum energy or cosmological constant as a possible candidate for dark energy. However, the energy scale related to the cosmological constant is $\sim 122$ orders of magnitude smaller than the energy of the vacuum originated at the Planck time (the cosmological constant problem) \cite{weinberg}. 

The following natural attempt to shed light on the problem is to consider a dynamical ``cosmological constant'' related to a scalar field whose slowly varying energy density is able to mimic an effective cosmological constant. That scalar field has been called quintessence \cite{steinhardt}, and a mechanism similar to the inflation but at lower energy scale has been suggested for the scalar field. The energy density of such a scalar field must remain sub-dominant at early stages and must start dominating the universe in the recent past (the cosmic coincidence problem), implying the requirement of specific evolution properties of the scalar field.

In the last few years, numerous quintessence models have been proposed (see {\it e.g.} \cite{steinhardt}, \cite{quint}). Most of these models assume minimally coupled scalar fields with different constraints and fine tuning of parameters for different types of potentials which model quintessence. In the context of General Relativity (GR) plus a minimally coupled scalar field, the scalar field and the potential of these models should be appropriately restricted by different observational data in order to get the ``right'' potential which could play the role of an effective cosmological constant. In that case, the scalar field potential can be constructed if  the luminosity distance (as a function of red-shift) can be known from observations. It is expected that the SNAP (Supernovae Acceleration Probe) satellite  will make accurate measurements of this parameter at a red-shift up to $z \sim 1.7$. This means that just the recent past of the universe will be probed and would be possible to construct just that corresponding part of the potential.

Minimally coupled self interacting scalar fields would be ruled out if the observational evidences point towards an equation of state with negative pressure for the dark energy. Furthermore, the violation of the identity ${dH^2(z)/ dz} \geq 3 \Omega_{m0}H_0(1+z)^2$ by models which assume minimally coupled scalar fields, and the characteristics of quintessence models mentioned above, enforce the idea of considering alternative theories where the scalar field is non-minimally coupled to gravity, like scalar-tensor theories (STT). 

Most of the attempts for using scalar-tensor theories have been done in the context of Brans-Dicke (BD) or some modified version of this theory \cite{bertolami}, \cite{sen}, \cite{sen-ses}, \cite{diego}. In the case of models in BD, some drawbacks have been found, {\it e.g.}, the difficulty for getting a transition from a decelerated expansion phase to an accelerated one (the solutions obtained in BD are always accelerated, in contradiction with the results of big-bang nucleosynthesis). A way to avoid this problem has been proposed by other authors \cite{diego}, using a modified version of BD theory by proposing the coupling parameter $\omega$ as a function of the scalar field $\phi$. A main difficulty of this last type of models have been the implication of a small value for $\omega$, in contradiction with bounds on $\omega$ obtained by solar system measurements.

Some other models have been proposed in the context of generalized scalar-tensor theories assuming different type of potentials ({\it see e.g.} \cite{esposito}). In some of these models a varying coupling function $\omega(\phi)$ has been suggested as a possible solution to alleviate discrepancies with observations and to be able to explain the late time behavior of the universe \cite{bartolo}, \cite{sen}, \cite{sen-ses}, \cite{diego}.

A variable coupling function $\omega(\phi)$ was first discussed by Nordtvedt \cite{nordtvedt} in the weak-field limit. Later on, Barker proposed a particular choice of $\omega(\phi)$ such that $dG/dt=0$ to first order in the weak-field limit \cite{barker}. More recent works \cite{barrow1} take into account that if $\omega$ varies, then it can increase with the cosmic time such that $\omega \to \infty$ and $\omega'\omega^{-3}$ $\to0$ as $t \to \infty$, and thus the weak-field observations at the present time agree with the predictions of GR, even though the theory may deviate from GR at very early cosmological times. In such works, the authors were able to get exact solutions for some specific cases of the equation of state, without putting attention in the deceleration parameter. A very extensive work has been done by Barrow and Parson \cite{bp} (see also \cite{navarro}) in the context of generalized scalar-tensor theories, which can approach GR in the weak-field limit at late time. They also propose three different functions for $\omega(\phi)$ and get asymptotic solutions at early and late cosmic times for different cases of an equation of state. They showed that a function $\omega(\phi)$, which fulfills the conditions in the weak field limit and that agrees with solar system measurements is not unique. This lack of uniqueness makes meaningless to look for analytic or numerical solutions of the field equations at all times, assuming a specific function for $\omega(\phi)$ (even if this function match the weak-field limit and agree with solar system measurements). 

However, taking into account the known conditions of a today flat and accelerated expanding universe, and the necessary transition from a decelerated radiation dominated epoch to an accelerated epoch today, we can analyze the function $\omega(\phi)$ as a cosmological parameter. On the other hand, as a local parameter we already know the values of $\omega(\phi)$ obtained from solar system measurements \cite{eubanks}, and as Scharre and Will \cite{will1} discuss, future observations of neutron stars spiraling into massive black holes by space-based laser interferometric detectors such as LISA, may place significant bounds on the scalar-tensor coupling parameter $\omega$. Then, we can have more accurate information to bound the function $\omega(\phi)$.

The aim of this work is to study the required conditions in generalized scalar-tensor theories that determine an accelerated expansion at the present epoch, a decelerated expansion in the past, appropriated limits on $\omega(\phi)$, and values of cosmological parameters consistent with measurements. 

We consider a generalized scalar-tensor theory with arbitrary $\omega(\phi)$ and $\Lambda(\phi)$, where this last function plays the role of an effective cosmological constant. In section 2, we solve our corresponding field equations for $\omega(\phi)$ and $\Lambda(\phi)$ as functions of the scalar field and the scale factor. We write those functions in terms of the deceleration and Hubble parameters. In section 3, we find expressions for the contributions of the scalar field, the effective cosmological constant, and the matter, to the density parameter in STT, assuming a flat universe. In section 4, we get a general expression of the deceleration parameter, and we show how it depends on the different contributions to the energy density parameter. In section 5, we discuss the conditions which $\omega(\phi)$ needs to satisfy in order to be consistent with solar system measurements. In section 6, we assume a power law of time for the scale factor and the scalar field, and we discuss the bounds on such power laws that determine an acelerated expanding model of the universe with cosmological parameters consistent with the values known from observations. We analyse the necessary conditions for a decelerated expanding model dominated by radiation. Finally, in section 7, we sumarize our results.

\section{Field Equations}

The action for a generalized scalar-tensor theory of gravitation is 
\cite{will}

\begin{equation}
\label{action}
S={1\over{16\pi G}}\int{d^4x\sqrt{-g}\,\bigg[\phi R-\phi^{-1}\omega(\phi) g^{\mu\nu} \partial_{\mu}\phi\partial_{\nu}\phi +2 \phi \Lambda (\phi)\bigg]}+S_{NG},
\end{equation}

\noindent where $R$ is the curvature scalar arising from the spacetime metric $g=\det g_{\mu\nu} $, $G$ is Newton's constant, and $S_{NG}$ is the action for the non-gravitational matter. The arbitrary functions $\omega(\phi)$ and $\Lambda(\phi)$ distinguish the different scalar-tensor theories of gravitation, $\Lambda(\phi)$ is a potential function and plays the role of an effective cosmological constant, $\omega(\phi)$ is the coupling function between the scalar field and gravity. The scalar field $\phi$ is a dynamical quantity, in contrast to Einstein's theory where $\phi \sim G^{-1}$ is a constant. Then, scalar-tensor gravity theories allow the value of the gravitational ``constant'' to vary, having as a particular case the Brans-Dicke theory for which $\omega$ is a constant. In this work we use the signature  $(-,+,+,+)$. 
 
The explicit field equations are

\begin{equation}
\label{field1}
G_{\mu \nu}={8 \pi T_{\mu \nu} \over {\phi}} + \Lambda (\phi) g_{\mu \nu} + 
\omega \phi^{-2}\left(\phi_{,\mu}\phi_{,\nu} - {1\over  2}g_{\mu\nu}\phi_{,\lambda}\phi^{,\lambda}\right)+
\phi^{-1}\left(\phi_{;\mu\nu}-g_{\mu\nu}\Box \phi\right),
\end{equation}

\begin{equation}
\label{field2}
\Box  \phi + {1\over 2}\phi_{,\lambda}\phi^{,\lambda}{d\over d\phi }\ln\left( {\omega (\phi) \over \phi}\right)+
{1\over 2}{\phi \over \omega (\phi)}\left[R+2{d\over d\phi }\left(\phi \Lambda(\phi)\right) \right]=0,
\end{equation}

\noindent where $G_{\mu \nu}$ is the Einstein tensor. The last  
equation can be rewritten as

\begin{equation}
\label{field3}
\Box \, \phi +{2\phi ^2 d\Lambda /d \phi -2\phi \Lambda(\phi) \over 3 + 2\omega (\phi)}=
{1\over 3+2\omega (\phi)}\left (\,8\pi T- {d\omega \over  d\phi} \phi_{,\mu} \phi ^{,\mu} \right ), 
\end{equation}

\noindent where $T=T_{\mu}^{\mu}$ is the trace of the stress-energy tensor. 
In a previous work \cite{coasting}, we have demonstrated that the divergenceless condition of the stress-energy matter tensor is satisfied if the field 
equation (\ref{field2}) is fulfilled, although our field equations are 
given by Eqs. (\ref{field1}) and (\ref{field3}).

\subsection{Friedmann universe}

We will study solutions to the previous field equations with time varying 
$G$, describing a homogeneous and isotropic Friedman-Robertson-Walker 
cosmological model, which line element in polar coordinates is given by  

\begin{equation}
\label{rw}
ds^2=-dt^2+a^2(t)\left [ {dr^2\over 1-k r^2}+
r^2\left(d\theta^2+\sin^2\theta \, \varphi^2\right) \right ].
\end{equation}

\noindent Here $(t,r,\theta, \varphi)$ are the polar coordinates, $k=-1, 0,  1$ are the 
values of the curvature parameter corresponding to open, flat or close 
cosmological models respectively, and $a(t)$ is the scale factor which characterize the expansion of the universe. With these considerations the field equations are given by 

\begin{equation}
\label{field11}
 3\Big({\dot a \over a}\Big)^2+{3k\over a^2}-\Lambda(\phi) - \frac{8\pi \rho}{\phi}  
-{\omega(\phi) \over 2}\Big({\dot \phi\over \phi}\Big)^2 +  3{\dot a\over a}{\dot \phi \over \phi}=0,
\end{equation}

\begin{equation}
\label{field12}  
2{{\ddot a}\over a}+\Big({\dot a\over a}\Big)^2+{k\over a^2} - \Lambda(\phi) + 
\frac{8\pi p}{\phi} + 
{\omega(\phi) \over 2}\Big({\dot\phi\over \phi}\Big)^2 + {{\ddot \phi}\over \phi} + 
2{\dot a \over a}{\dot \phi\over \phi}=0,
\end{equation}

\begin{equation}
\label{field13}
 \left[{{\ddot \phi}\over \phi}+3{\dot a\over a}{\dot\phi\over \phi}\right]\Big(3+2\omega(\phi)\Big) - 2\left({\Lambda(\phi)-\phi {d\Lambda \over d\phi}}\right) - {8\pi \over \phi}(\rho-3 p) + 
{d\omega(\phi) \over d\phi} {{\dot \phi}^2 \over \phi}=0,
\end{equation}

\noindent and the energy conservation equation 

\begin{equation}
\label{den-eq}
{\dot \rho} + 3 {\dot a \over a}(\rho + p)=0.
\end{equation}
                                                                       
\noindent Here $a(t)$, $\phi(t)$, $\omega(\phi)$, $\Lambda(\phi)$, $p(t)$ and $\rho(t)$ are our variables and the derivatives respect the cosmic time $t$ are denoted by over-dots. Of the set of field equations (\ref{field11})-(\ref{den-eq}), three of them are independent and we have to deal with six unknowns, so we need some assumptions to match the number of our independent field equations and variables. We will deal with our set of field equations without specifying any assumption for the moment, to get a general functional form of the variables which we are interested in.

\noindent We will assume that the universe contains a perfect fluid with a barotropic state equation $p=(\gamma-1)\rho$, where $\gamma$ is a constant lying in the range $ 0 \leq \gamma \leq 2 $. Then from Eq. (\ref{den-eq}), we have

\begin{equation}
{\dot \rho} + 3 \gamma {\dot a \over a} \rho=0.
\end{equation}

\noindent This equation has the following solution 

\begin{equation}
\label{density}
\rho = \rho_1 a^{-3 \gamma},
\end{equation}

\noindent where $\rho_1$ is an integration constant. Thus, our field equations will be (\ref{field11})-(\ref{field13}) where $\rho$ is given by Eq. (\ref{density}).

\subsection{Reduction of the system of equations}

We can combine equations (\ref{field11})-(\ref{field13}) to reduce our set of field equations. Thus we get from Eq. (\ref{field11})

\begin{equation}
\label{lambda-phi}
\Lambda(\phi)=3{{\dot a}^2\over a^2} + 3 {\dot a\over a}{\dot \phi \over \phi}-
{\omega(\phi)\over 2}{{\dot \phi} ^2 \over \phi^2}-{8\pi \rho \over \phi} + {3k \over a^2}.
\end{equation}

\noindent Using this last equation and its derivatives in Eqs. (\ref{field12})-(\ref{field13}), we get respectively

\begin{equation}
\label{field21}
{\ddot \phi \over \phi} + 2 {\ddot a \over a} + \omega(\phi) {\dot \phi^2 \over \phi^2} - 2 {\dot a^2 \over a^2} - {\dot a \over a}{\dot \phi \over \phi} + {8\pi \rho \over \phi} \gamma - {2k \over a^2} = 0,
\end{equation}

\begin{equation}
\label{field22}
\left( {\dot a \over a} + {1\over 2}{\dot \phi \over \phi}\right)\left[ {\ddot \phi \over \phi} + 2{\ddot a \over a} + \omega(\phi) {\dot \phi^2 \over \phi^2} - 2 {\dot a^2\over a^2} - {\dot a \over a}{\dot \phi \over \phi} + {8\pi\rho \over \phi}\gamma -{2k \over a^2}\right]=0.
\end{equation}

\noindent The term between square brackets in the last equation, is equal to the left hand side of Eq. (\ref{field21}). Then our set of 3 field equations has been reduced to a single equation for 3 variables

\begin{equation}
\label{fiel}
{\ddot \phi \over \phi} + 2 {\ddot a \over a} + \omega(\phi) {\dot \phi^2\over \phi^2} - 2 {\dot a^2\over a^2} - {\dot a \over a}{\dot \phi \over \phi} + {8\pi \rho \over \phi} \gamma - {2k \over a^2}=0.
\end{equation}  

\noindent We can write this last equation for $\omega(\phi)$ as follows

\begin{equation}
\label{omega-phi-a}
\omega(\phi)=\left( {\dot \phi \over \phi} \right)^{-2} \left[ -{\ddot \phi \over \phi} -2 {\ddot a \over a} + 2 {\dot a^2 \over a^2} + {\dot a \over a}{\dot \phi \over \phi}-{8\pi \rho \over \phi} \gamma + {2k \over a^2} \right].
\end{equation}

\noindent Our set of field equations are satisfied for $\omega(\phi)$ given in Eq. (\ref{omega-phi-a}) and $\Lambda(\phi)$ given by Eq. (\ref{lambda-phi}), independently of the form of the functions $a(t)$ and $\phi(t)$. We can write Eq. (\ref{lambda-phi}) just in terms of $a(t)$ and $\phi(t)$ as follows 

\begin{equation}
\label{lambda-phi-a}
\Lambda(\phi)={\ddot a \over a} + {1\over 2}{\ddot \phi \over \phi}+2 {\dot a^2 \over a^2}+{5\over 2}{\dot a \over a}{\dot \phi \over \phi} + {8\pi \rho \over \phi}\left( {\gamma \over 2} -1\right) + {2k \over a^2}.
\end{equation}

\noindent It is convenient, for the purpose of the present work, to write $\omega(\phi)$ and $\Lambda(\phi)$ as functions of the deceleration parameter defined, as usually, by $q=-{{\ddot a}a \over {\dot a^2}}$, thus we have

\begin{equation}
\label{omega-q}
\omega(\phi)=\left( {\dot \phi \over \phi}\right)^{-2}\left[ -{\ddot \phi \over \phi} + 2H^2 (1+q) + H{\dot \phi \over \phi} - {8\pi \rho \over \phi}\gamma + {2k \over a^2} \right],
\end{equation}

\begin{equation}
\label{lambda-q}
\Lambda(\phi)={1\over 2}{\ddot \phi \over \phi} + H^2(2-q) + {5\over 2}H{\dot \phi \over \phi}+{8\pi \rho \over \phi}\left( {\gamma \over 2}-1\right)+{2k \over a^2}.
\end{equation}   

\noindent The field equations of a general scalar-tensor theory, Eqs. (\ref{field11})-(\ref{den-eq}), are satisfied if the coupling function $\omega(\phi)$ and the effective cosmological constant $\Lambda(\phi)$ are given by the previous two equations. 

It has been assumed in the standard literature some functional form for $\omega(\phi)$ and $\Lambda(\phi)$ to close the system of Equations (\ref{field11})-(\ref{den-eq}). Instead of doing the same, we have obtained $\omega(\phi)$ and $\Lambda(\phi)$ as functions of $a(t)$ and $\phi(t)$ which can be proposed and restricted according with observational results like $H_0$, $q_0$, $\Omega_{m0}$, {\it etc.}

\section{The density parameter}

We can obtain an expression for the contributions to the density parameter from the matter and the scalar field without specifying the explicit functions of $a(t)$ and $\phi(t)$. We have, from Eq. (\ref{field11}) for a flat case

\begin{equation}
\label{field11-Om}
1={8\pi G \over 3H^2}\left[ {\Lambda(\phi) \over 8\pi G} + {\rho \over G }{1 \over \phi } + {\omega(\phi) \over 16\pi G}{\dot \phi^2 \over \phi^2}-{3H \over 8\pi G}{\dot \phi \over \phi} \right].
\end{equation}

\noindent Defining \cite{lambda1}

\begin{equation}
\label{omega-m}
\Omega_m = {8\pi \rho \over 3H^2}{1\over \phi},
\end{equation}

\begin{equation}
\label{omega-lambda}
\Omega_{\Lambda} = {\Lambda(\phi) \over 3H^2},
\end{equation}

\begin{equation}
\label{omega-phi}
\Omega_{\phi} = {1 \over 3H^2}\left[ {1\over 2}\omega(\phi){\dot \phi^2 \over \phi^2}-3H{\dot \phi \over \phi} \right], 
\end{equation}

\noindent we have 

\begin{equation}
\label{sum-omegas}
1 = \Omega_m + \Omega_{\Lambda} + \Omega_{\phi}.
\end{equation}

\noindent It is important to emphasize that $\Omega_m$, $\Omega_{\Lambda}$ and $\Omega_{\phi}$ are here 
mere parameters which we have defined in analogy with theirs definitions in GR.

\noindent Using Eqs. (\ref{omega-q})-(\ref{lambda-q}), we write

\begin{equation}
\label{omega-lamb-q}
\Omega_{\Lambda}={1\over 3H^2}\left[ {1\over 2}{\ddot \phi \over \phi} + H^2(2-q) + {5\over 2}H{\dot \phi \over \phi} + {8\pi \rho \over \phi}\left( {\gamma \over 2}-1 \right) \right],  
\end{equation}

\begin{equation}
\label{omega-phi-q}
\Omega_{\phi}=-{1 \over H}{\dot \phi \over \phi}+ {1\over 6H^2}\left[ -{\ddot \phi \over \phi} + 2H^2(1+q) + H{\dot \phi \over \phi} - {8\pi \rho \over \phi}\gamma \right],
\end{equation}

\noindent where $\rho$ is given by Eq. (\ref{density}). Here $a(t)$ and $\phi(t)$ have not been specified yet .

\section{The deceleration parameter}

As in the previous sections, we write an expression for the deceleration parameter in a similar way as it has been done in GR. We start with the difference between the Friedmann equation for the present theory, Eq. (\ref{field11}), and Eq. (\ref{field12}), to obtain, like in GR, an equation for the acceleration

\begin{equation}
\label{q-gen}
{{\ddot a} \over a}=-{4\pi \over 3 \phi}(\rho + 3p)+{1\over 3}\left( \Lambda(\phi)-\omega(\phi){\dot \phi^2 \over \phi^2}\right) -{1\over 2}\left( {\ddot \phi \over \phi} + H {\dot \phi \over \phi }\right),
\end{equation}

\noindent using the definition of the deceleration parameter $q=-{{\ddot a}a \over {\dot a^2}}$ in the previous equation, as well as the definitions given by Eqs. (\ref{omega-m})-(\ref{omega-phi}), we get

\begin{equation}
\label{q-omegas-p}
q={1\over 2}\Omega_{m}\left(1+3{p\over \rho}\right)-\Omega_{\Lambda}+2\Omega_{\phi}+{1\over 2H^2}\left( {\ddot \phi \over \phi}+5H{\dot \phi \over \phi}\right).
\end{equation}

\noindent We note from this equation that in the limit of GR we get the well known result $q={1\over 2}\Omega_{m}(1+3{p\over \rho})$, from where it is clear that $p<-\rho/3$ is the condition required for an accelerated expanding model. In our case, we note from Eq. (\ref{q-omegas-p}) that $q<0$ depends also on the values of $\Omega_\Lambda$, $\Omega_\phi$ and the scalar field $\phi$.

\noindent As we have mentioned in Section 2.1, we assume a perfect fluid with a barotropic equation of state $p=(\gamma -1)\rho$, thus taking this into account and rearranging terms in the previous equation, we obtain 

\begin{equation}
\label{q0-omegas}
q_0={1\over 2}\Omega_{m0}-{1\over 2}\left[ 3(1-\gamma)_0+(3\gamma-1)_0\Omega_{\Lambda0}+(3\gamma-7)_0\Omega_{\phi0}-{1\over H_0^2}\left( {\ddot \phi \over \phi}+5H{\dot \phi \over \phi} \right)_0 \right],
\end{equation}

\noindent where the zero subindex indicates the corresponding values today. From this last expression for $q_0$, it is clear that $q_0<0$ implies

\begin{equation}
\Omega_{m0}< 3(1-\gamma_0)+(3\gamma_0-1)\Omega_{\Lambda0}+(3\gamma_0-7)\Omega_{\phi0}-{1\over H_0^2}\Bigg( {\ddot \phi \over \phi}+5H{\dot \phi \over \phi}\Bigg)_0,
\end{equation}

\noindent {\it i.e.}, according with the present theory, $q_0$ is negative no just due to the state equation but also to contributions of the scalar field to the density parameter.

\noindent From the theoretical point of view, it is expected that accelerated expansion is a recent phenomena \cite{turner-riess}. From the observational point of view, there are some evidences of a decelerated phase, like observations of SN 1997ff, a supernova type Ia at $z\sim 1.7$ \cite{riess2001} and some other examples of this kind of stars at $z\sim 1.2$ \cite{tonry}. This suggest a transition from a decelerated expansion epoch to an accelerated one, which takes place when $q=0$. Thus according with this theory, from Eq. (\ref{q-omegas-p}) at that specific epoch when the transition took place, we get the condition

\begin{equation}
\label{omegas-r}
\Omega_m(z_{tr})=3(1-\gamma_{tr})+(3\gamma_{tr}-1)\Omega_{\Lambda}(z_{tr})+(3\gamma_{tr}-7)\Omega_\phi(z_{tr}){1\over H^2(z_{tr}) }\left( {\ddot \phi \over \phi}+5H{\dot \phi \over \phi}\right)_{tr},
\end{equation}

\noindent where $\Omega_m$, $\Omega_\Lambda$, $\Omega_\phi$, and $H$ have been written as functions of the red-shift at the epoch when the transition took place. The subindex $tr$ refers to the value of the corresponding parameters at the transition epoch. Here again $a(t)$ and $\phi(t)$ have not been specified yet.

\section{The coupling function $\omega(\phi)$}

A coupling arbitrary function $\omega(\phi)$ enlarges the possibilities for variations of $G$. In the weak field limit, Nordtvedt \cite{nordtvedt} found an expression for the observed value of the gravitational constant 

\begin{equation}
G(t)=\phi^{-1}\left({{4+2\omega(\phi)}\over {3+2\omega(\phi)}}\right),
\end{equation}

\noindent and Barker \cite{barker}  found that, with a particular choice of $\omega(\phi)$, is possible to have $\dot{G}=0$ to first order in the weak field limit. 

\noindent The known observational limits on $\omega$ and $\dot G$ are a mixture of weak field test of gravitation in the solar system, and of cosmological limits. For theories with $\omega$ constant, the possibility of variation of $G$ is very small. However if $\omega(\phi)$ varies, then it can increase with cosmic time, such that in the weak field limit, the theory tend to GR

\begin{equation}
\label{omega-cond-bp}
\omega(\phi) \longrightarrow \infty \, \, \, \, {\rm and} \, \, \, \, {d\omega(\phi)\over d\phi} {1\over \omega^3}\longrightarrow 0 \, \, \, \, \, \, \, \, \, \, {\rm as} \, \, \, \phi \longrightarrow \phi_0 \, \, \, ({\rm or} \,  t \longrightarrow \infty),
\end{equation} 

\noindent where $\phi_0$ is the scalar field, evaluated at the present epoch $t=t_0$. Thus, weak field observations at the present time are consistent with GR's predictions. As Barrow and Parson discussed \cite{bp}, there are many functions $\omega(\phi)$ which satisfy the first condition ($\omega(\phi) \longrightarrow \infty $), but not the second one (${d\omega(\phi)\over d\phi} {1\over \omega^3}\longrightarrow 0 $). The same authors showed that there are more than one function $\omega(\phi)$ which fulfills both required limit conditions. Therefore, it will not be very meaningful to assume a function for $\omega(\phi)$ which satisfies the required limit conditions, and look for cosmological solutions at all times, because such a $\omega(\phi)$ is not unique and the ``true'' function can be very different.

\noindent Now, a new observational fact has been added to the cosmological scenario, namely the recent discovery of the accelerated expansion of the universe \cite{perlmutter}. Then, any theoretical cosmological model should take it into account as an important requirement for a late epoch of any cosmological model.
 A decelerated expansion is also expected in an early epoch dominated by radiation, after an inflationary period at the beginning of the universe. 
 
\noindent In order to look for answers to the today accelerated expansion of the universe, some authors have studied cosmological models in Brans-Dicke theory or some modified version of this theory (see for example \cite{sen-ses}, \cite{diego}, \cite{bartolo}). A small negative value of $\omega$ at $t \to \infty $ has been a recurrent result in that studies, in contradiction with values obtained from solar system measurements. This results suggest that, a possible solution of this problem, is to consider a non-constant coupling function $\omega(\phi)$. Thus, the value of such a function can change with the cosmic time and, in the limit $t \to \infty $, it could agree with local measured values.

\noindent As we have pointed out before, it is not enough to assume an arbitrary $\omega(\phi)$, even if it fulfills the required limit conditions. We need to look for additional consistent conditions on $\omega(\phi)$ in order to satisfy the weak field limit condition, as well as a present accelerated expansion after a transition from a decelerated epoch dominated by radiation, and the correct values of the cosmological parameters consistent with observational results.

\section{Power law evolution}


\begin{figure}[!hbp]
\begin{center}
\includegraphics[width=220pt]{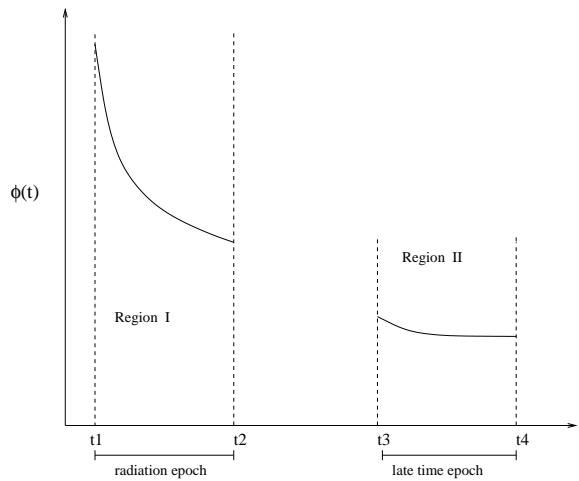}{}

\end{center}

\vspace*{8pt}

\caption[]{Qualitative graphics of $\phi(t)$ {\it vs} $t$ from Eq. (\ref{a-phi}). Region II corresponds to a late time epoch and small values of $\beta$ ($\beta \sim 0$), region I corresponds to bigger values of $\beta$ at the radiation dominated epoch.}

\end{figure}


We have started with a non-close set of field equations which give us the freedom to assume a functional form for $a(t)$ and $\phi(t)$. Let us assume for simplicity that both, the scale factor and the scalar field evolve as power law functions of time \cite{sen-ses} 

\begin{equation}
\label{a-phi}
a(t)=a_0 \left( {t \over t_0} \right)^{\alpha}, \, \, \, \, \, \, \, \, \, \, 
\phi(t)= \phi_0 \left( {t \over t_0 } \right)^{\beta},
\end{equation}    

\noindent where $\alpha$ and $\beta$ are constants and $a_0$ and $\phi_0$ are the values of the parameters at the present epoch, $t=t_0$.

One of the main characteristics of the present theory is that $\phi(t)$ (as well as $a(t)$) is a dynamical parameter, from which we are trying to explain the accelerated expansion concluded from observations. However, a unique value for the constants $\alpha$ and $\beta$, in the assumed functions (\ref{a-phi}), cannot describe the expected behavior of the universe at early and late times, and also satisfy the weak field limits (unless $\alpha$ and $\beta$ would be also functions of time). If we want to use equations (\ref{a-phi}), it is necessary to consider at least two regions of these functions, corresponding to two different values of $\alpha$ and $\beta$. One region should consider small values of $\beta$ such that as $t \to t_0$ we can get $\beta \sim 0$ (we will show that this is a condition for satisfying the weak field limits). The second region should consider values of $\big |\beta\big | \neq 0$. In a similar way, we expect that $\alpha$ get different values at late times than those at early times.

The ideal case, of course, is to assume more general functions for $a(t)$ and $\phi(t)$ which could describe continuously the behavior of the universe at any time, but this is not an easy task with the information available now. 

Taking the previous assumptions into account, we will consider two regions of the functions (\ref{a-phi}) (see figure 1), one for small values of $\beta$ at late times $t \to t_0$, and another for larger values of $\big |\beta\big |$ at early times ({\it e.g.} $t \sim t_{radiation}$). These assumptions on $\beta$ necessarily imply different values of $\alpha$ at the two considered regions. Those values should be obtained as required conditions to agree with values of cosmological parameters, as we will discuss in the following sections.

\bigskip

From the assumed function for $a(t)$, we get immediately an expression for the deceleration parameter

\begin{equation}
\label{q}
q={1\over \alpha}-1,
\end{equation}

\noindent thus, an accelerated expansion implies $\alpha > 1$.

\noindent Once we known the functional form of $a(t)$ and $\phi(t)$, we can write Eqs. (\ref{omega-q}) and (\ref{lambda-q}) as explicit functions of the cosmic time:

\begin{equation}
\label{omega-t}
\omega(t)=-1 + {1\over \beta} + {\alpha \over \beta} + 2{\alpha \over \beta^2}- {1\over \beta^2}{8\pi \rho_1 \over \phi_0 a_0^{3\gamma}}\gamma t_0^{3\gamma \alpha + \beta}t^{2-3\gamma \alpha -\beta}+{1\over \beta^2}{2k \over a_0^2}t_0^{2\alpha}t^{2(1-\alpha)},
\end{equation}

\begin{eqnarray}
\label{lambda-t}
\Lambda(t)=t^{-2}\left[ {\beta^2 \over 2}- {\beta \over 2}+3\alpha^2-\alpha+{5\over 2}\alpha \beta \right]+{8\pi \rho_1 \over \phi_0 a_0^{3\gamma}} t_0^{3\gamma \alpha + \beta} t^{-3\gamma \alpha -\beta} \left( {\gamma \over 2}-1 \right) \nonumber \\
+ {2k \over a_0^2} t_0^{2\alpha} t^{-2\alpha}.
\end{eqnarray}

\noindent We also get an expression for the following quantity, which at the weak field limit needs to satisfy the limit condition (see Eq. \ref{omega-cond-bp})

\begin{eqnarray}
\label{omega-d}
{\dot \omega \over \omega^3}{1\over \dot \phi}=&& \beta^3 {t_0^{\beta}\over \phi_0}t^{2-2\alpha - \beta}\left[ (\beta + 3\gamma \alpha -2){8\pi \rho_1 \over \phi_0 a_0^{3\gamma}}\gamma t_0^{3\gamma \alpha + \beta}t^{2\alpha-3\gamma \alpha - \beta}+(1-\alpha){4k \over a_0^2}t_0^{2\alpha}\right]  \nonumber \\
&& {\times} \left[ 1-\beta^2 + \alpha \beta + 2\alpha - {8\pi \rho_1 \over \phi_0 a_0^{3\gamma}}\gamma t_0^{3\gamma \alpha + \beta}t^{2-3\gamma \alpha - \beta}+{2k \over a_0^2}t_0^{2\alpha}t^{2(1-\alpha)} \right]^{-3}.
\end{eqnarray}

\noindent At the present time ($t=t_0$), for a flat case (k=0), the three previous equations can be written as follows

\begin{equation}
\label{omega-0t}
\omega(t_0)=-1 + {1\over \beta} + {\alpha \over \beta} + 2{\alpha \over \beta^2}- {1\over \beta^2}{8\pi \rho_1 \over \phi_0 a_0^{3\gamma}}\gamma t_0^2,
\end{equation}

\begin{equation}
\label{lambda-t0}
\Lambda(t_0)=t_0^{-2}\left[ {\beta^2 \over 2}- {\beta \over 2}+3\alpha^2-\alpha+{5\over 2}\alpha \beta \right]+{8\pi \rho_1 \over \phi_0 a_0^{3\gamma}} \left( {\gamma \over 2}-1 \right),
\end{equation}

\begin{eqnarray}
\label{omega-d0}
{\dot \omega \over \omega^3}{1\over \dot \phi}|_{t=t_0}=\beta^3 {1\over \phi_0}t_0^{2-2\alpha}\left[ (\beta + 3\gamma \alpha -2){8\pi \rho_1 \over \phi_0 a_0^{3\gamma}}\gamma t_0^{2\alpha}\right] \nonumber \\
\times \left[ 1-\beta^2 + \alpha \beta + 2\alpha - {8\pi \rho_1 \over \phi_0 a_0^{3\gamma}}\gamma t_0^{2}\right]^{-3}.
\end{eqnarray}

\noindent From the discussion of section 5, we know that in order to satisfy the weak field limits, we need to satisfy Eq. (\ref{omega-cond-bp}), at $t=t_0$, {\it i.e.}, $\omega(t_0) \longrightarrow \infty$ and ${\dot \omega \over \omega^3}{1\over \dot \phi}|_{t=t_0}=0$. From inspection of Eqs. (\ref{omega-0t}) and (\ref{omega-d0}), we note that the conditions of weak field limit are satisfied if $\beta =0$. However, this value of $\beta$ cannot be used for another value of the cosmic time different than $t_0$, because from Eq. (\ref{a-phi}), it would imply $\phi(t)=\phi_0=constant$ for all times, which is not correct. This is the reason why we are assuming two regions of validity for Eqs. (\ref{a-phi}).

\subsection{Region of small values of $\beta$. Conditions for late time acceleration} 

We have seen that in a region of $t \sim t_0$ and small values of $\beta$ (region II in Fig. I), the conditions of weak field limits are satisfied. In this same region, the effective cosmological constant given by Eq. (\ref{lambda-t0}) is written as 

\begin{equation}
\label{lambda-t0-b0}
\Lambda(t_0)=t_0^{-2}\left[3\alpha^2-\alpha\right]+{8\pi \rho_1 \over \phi_0 a_0^{3\gamma}} \left( {\gamma \over 2}-1 \right).
\end{equation}

\noindent We will consider two examples of an equation of state at the present time, $t=t_0$: a universe dominated by cold matter $\gamma=1$, and a universe dominated by ``false vacuum'', $\gamma=0$ (assuming in both cases $k=0$). We also discuss the general case of an arbitrary state equation.

\begin{itemize}

\item Matter dominated model, $\gamma=1$.

For this case of the equation of state, we have from Eq. (\ref{lambda-t0-b0})

\begin{equation}
\label{lambda-0}
\Lambda(t_0)=H_0^{2}(3\alpha^2-\alpha)-{8\pi \rho_1 \over 2\phi_0}a_0^{-3},
\end{equation}

\noindent $\Lambda(t_0)$ has a constant value depending on $\alpha$. The only restriction on $\alpha$ is given by Eq. (\ref{q}), i.e. $\alpha>1$ for accelerated expansion. Let us assume $q_0=-1/4$ $\Longrightarrow $ $\alpha =4/3$ from Eq.(\ref{q}) (similar conditions found Bertolami and Martin \cite{bertolami} in their model and Sen and Sethi \cite{sen} found that the best fit value with SN Ia data is $\alpha \sim 5/4 $, both values of $\alpha$ give $q_0 \sim - 0.2$). Also let us assume $H_0=65 \, {\rm km \, s^{-1} \, Mpc^{-1}}$, thus we have from Eq. (\ref{lambda-0}) 

\begin{equation}
\label{lambda-0-max}
\Lambda(t_0) \approx 1.76 \times 10^{-35} s^{-2}-{8\pi \rho_1 \over 2\phi_0}a_0^{-3}.
\end{equation}

\noindent According with the measurements, $\Lambda_0\sim 10^{-35}\, {\rm s^{-2}}$, then, for the previous equation to be of the same order of the corresponding measured value, it is necessary that ${8\pi \rho_1 \over 2\phi_0}a_0^{-3} \leq 7.6 {\rm x}10^{-36}$.

The parameters defined by Eqs. (\ref{omega-m}), (\ref{omega-lamb-q}) and (\ref{omega-phi-q}), for the same region, {\it i.e.}, $t=t_0$, $\beta=0$, $\alpha=4/3$ ($q_0=-1/4$) and for $\gamma=1$, are given by

\begin{equation}
\label{omega-m-00}
\Omega_{m0}= {3\over 8}H_0^{-2}{8\pi \rho_{1} a_0^{-3} \over 2 \phi_0},
\end{equation}

\begin{equation}
\label{omega-lambda-00}
\Omega_{\Lambda 0}={3 \over 4}-{3\over 16}{8\pi \rho_{1} a_0^{-3} \over 2\phi_0}H_0^{-2},
\end{equation}

\begin{equation}
\label{omega-phi-00}
\Omega_{\phi 0}={3\over 16}\left[ {4\over 3} - {8\pi \rho_{1} a_0^{-3} \over 2\phi_0}H_0^{-2} \right].
\end{equation}

\noindent We found from Eq. (\ref{lambda-0-max}) that for $\Lambda(t_0)$ to agree with the observations, the maximum allowed value for $8\pi \rho_1 (2\phi_0)^{-1}a_0^{-3}$ is $ \sim 7.6 \, \times \, 10^{-36}\, \, {\rm s^{-2}}$, then using this value in Eqs. (\ref{omega-m-00})-(\ref{omega-phi-00}) we get the values for our parameters $\Omega_{m0} \approx 0.64$, and $\Omega_{\Lambda0}+\Omega_{\phi0} \approx 0.35$. 

\noindent These values are clearly different than the values of the density parameters estimated from observations. $\Omega_m =0.33 \pm 0.04$ has been obtained from measurements of CMB anisotropies of bulk flows, and of the baryonic fraction in clusters \cite{bernadis},\cite{turner01}, where the energy density of the dark energy has been estimated as $\Omega_x$=$\Omega_\Lambda = 0.66 \pm 0.1$. However, it is important to take into account that these measured values can be strongly dependent on the assumed cosmological model for theirs calculation. We note that  we have assumed in our model a specific value for the deceleration parameter $q_0=-1/4$, as well as a universe dominated by matter. 

\item ``False vacuum'' dominated model, $\gamma=0$.

For a ``false vacuum'' dominated universe, $\gamma=0$ in the equation of state, and as in the previous case, restricting our analysis to the late time region: $\alpha=4/3$ and $\beta=0$ for $t=t_0$, and  $8\pi \rho_1 (\phi_0)^{-1}\approx 7.6 \, {\rm x} \, 10^{-36}\, \, {\rm s^{-2}}$, we have

\begin{eqnarray}
\Omega_{m0}={3\over 16}{8\pi \rho_1 \over \phi_0}H_0^{-2} \sim 0.323, \nonumber \\
\Omega_{\Lambda0}+\Omega_{\phi0}={3\over 4}-{3\over 16}{8\pi \rho_1 \over \phi_0}H_0^{-2}+{1\over 4} 
\sim 0.675,
\end{eqnarray}
 
\noindent both values of our parameters are in agreement with the measurements of $\Omega_{m0}$ and $\Omega_{\Lambda_0}$.

\noindent In spite of this results, which apparently favors a model with $p=-\rho$, we cannot conclude that the theory which we are analyzing here, predicts a model dominated by ``false vacuum'', because we have assumed a specific value of $\alpha$ (or $q_0$), which can be different from the ``true'' value if the actual evolution of the scale factor is given by Eq. (\ref{a-phi}).

\item Undetermined perfect fluid.

Let us consider the general case of an arbitrary perfect fluid with a barotropic state equation  $p_0=(\gamma-1)\rho_0$, where the subindex indicate the corresponding values at the present time . We have from Eqs. (\ref{omega-m}) and (\ref{a-phi})

\begin{equation}
\label{8pi-gen}
{8\pi \rho_1 \over \phi_0}a_0^{-3\gamma_0}=3\alpha^2\Omega_{m0}H_0^2.
\end{equation}

\noindent Using this equation in Eq. (\ref{lambda-t0-b0}) (with $\gamma$ undetermined), and rearranging terms, we have

\begin{equation}
\label{alpha-gen}
\alpha^2\left[3H_0^2+3\Omega_{m0}H_0^2\left({\gamma \over 2}-1\right)\right]- \alpha H_0^2- \Lambda(t_0)=0.  
\end{equation}

\noindent The solution for $\alpha$ clearly depends on $\gamma$ and {\it vice versa}

\begin{equation}
\label{alpha-sol-gen}
\alpha={ 1 \pm \sqrt{1+12 \Lambda(t_0)H_0^{-2}\left[1+\Omega_{mo}\left({\gamma \over 2}-1\right)\right]} \over 6\left[1+\Omega_{mo}\left({\gamma \over 2}-1\right)\right]}.
\end{equation}

\noindent Thus, assuming a value of one of this constants, $\alpha$ or $\gamma$, clearly determines the value of the other one. Let assume $\gamma=0$ (false vacuum), for example, and using in the previous equation the values $H_0=65 \, {\rm km \, s^{-1} \, Mpc^{-1}}$,  $\Lambda_0\sim 10^{-35}\, {\rm s^{-2}}$, and $\Omega_{m0}=0.33$, we get $\alpha_1=1.339$ and $\alpha_2=-0.842$. The requirement for an accelerated expansion ($q<0$) is $\alpha>1$ (see Eq. ({\ref{q}})), thus we chose the value $\alpha=1.339 \sim 4/3$, precisely the value of $\alpha$ which we have assumed in the previous examples of mater and ``false vacuum'' dominated models, and which explains why the results seems to favor a ``false vacuum'' dominated model. 

\noindent As another example, let us assume $\gamma=1$ (dust model) and by calculating $\alpha$ from Eq. (\ref{alpha-sol-gen}), we get $\alpha_1\sim 1.171$ and $\alpha_2 \sim -0.772$. Thus, for a model with a dust state equation and with accelerated expansion, which match correctly the measured values of $\Lambda_0$, $H_0$, and $\Omega_{m0}$, it is necessary to have $\alpha \sim 1.171  \sim 7/6$, or $q_0 \sim -0.146$. 

\end{itemize}

\noindent From the proposed function for the scale factor, Eq. (\ref{a-phi}), we have $t=\alpha/H$. For an uniform expansion $\alpha=1$ $\Longrightarrow$ $t_0=4.7 \times 10^{17}\,{\rm s}\, \approx  15 \, {\rm Gyr}$ which is a quit large age, if the accepted age of the universe is $\sim 13.4 \pm 1.6 \, \, {\rm Gyr}$ \cite{line}, and hence according with this model, the larger the value of $\alpha$ or the acceleration, the older the universe is. In particular for $\alpha=4/3$,  we get $t_0=6.3 \times 10^{17}\,{\rm s}\, \approx  20 \, \, {\rm Gyr}$, which is a large value. For $\alpha \sim 7/6$ we get $t_0=5.57 \times 10^{17}\,{\rm s}\, \approx  17.6 \, \, {\rm Gyr}$, which is still a large value.

\subsection{Region of large values of $\big |\beta\big |$. Conditions for early time deceleration}


\begin{figure}[hbp]
\begin{center}

\includegraphics[width=220pt]{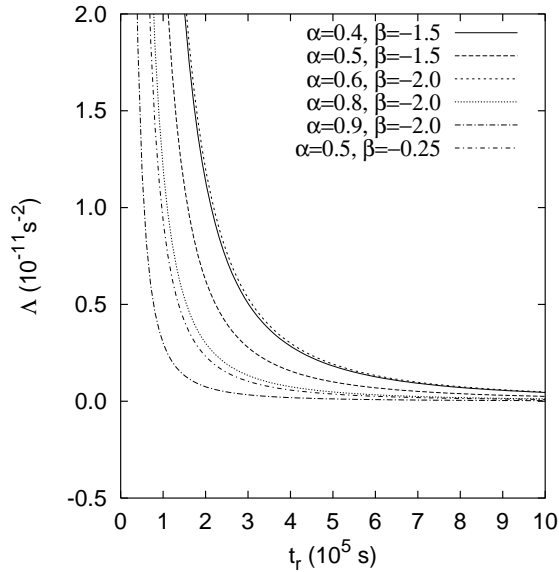}{}

\end{center}

\vspace*{8pt}

\caption[]{$\Lambda$ {\it vs} $t_r$ from Eq. (\ref{lambda-r}) for different values of $\alpha$ and $\beta$ which satisfy the restrictions $4\alpha+\beta>0$, $\alpha<1$, and $\beta<0$. We see that $\Lambda$ decreases for all values of $t_r$ in the plotted interval. We have assumed $H_0=65 \, {\rm km\, \, s^{-1}\, \, Mpc^{-1}}$.}

\end{figure}


\begin{figure}[hbp]
\begin{center}

\includegraphics[width=220pts]{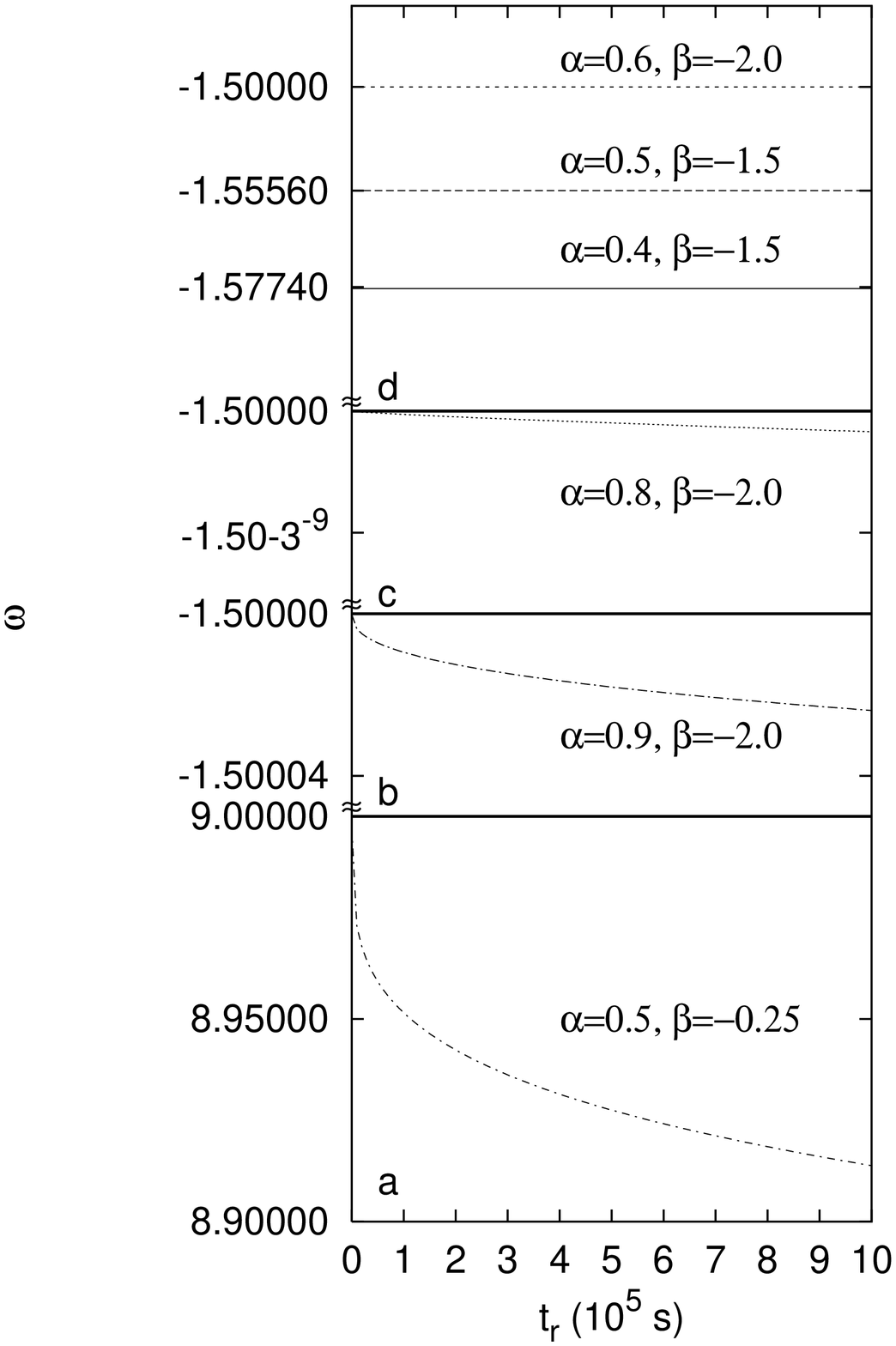}{}

\end{center}

\vspace*{8pt}

\caption[]{$\omega$ {\it vs} $t_r$ from Eq.(\ref{omega-r}) for different values of $\alpha$ and $\beta$ which satisfy the restrictions $4\alpha+\beta>0$, $\alpha<1$, and $\beta<0$. Note the scale of the {\it y} axis, thus we can consider that $\omega$ plotted in panels (b)-(d), remains practically constant in the interval from $1$ to $10^6 \, {\rm s}$. Panel (a) shows that for the corresponding values of $\alpha$ and $\beta$, $\omega$ is a slow decreasing function.}

\end{figure}


The big-bang nucleosynthesis model has successfully explained the abundances of light elements \cite{burles}, and recent measurements of baryon density by CMB measurements, have confirmed the predictions of the big-bang nucleosynthesis model \cite{pryke}, \cite{netterfield}. This confirmation is a strong evidence that the universe was dominated by radiation, and expanding in a decelerated way, seconds after the big-bang. The discovery of high red-shift SNe Ia SN 1997ff at $z \sim 1.7$ \cite{riess2001}, provides evidence for an epoch of slow expansion. This is true in GR if the dark energy is a cosmological constant, as has been pointed out by Turner and Riess \cite{turner-riess}. In this subsection we want to discuss under which conditions of the present theory, is possible a period of decelerated expansion, for a model dominated by radiation. 

From Eqs. (\ref{omega-t})-(\ref{lambda-t}) with $\gamma =4/3$ for the radiation epoch (region I in Fig. 1), we have 

\begin{equation}
\label{lambda-r}
\Lambda(t_r)=t_r^{-2}\left[ {\beta^2 \over 2}-{\beta \over 2}+3\alpha^2 -\alpha +{5\over 2}\alpha \beta \right] - {8\pi \rho_1 \over 3 \phi_0 a_0^4}t_0^{4\alpha +\beta} t_r^{-4\alpha -\beta}+{2k \over a_0^2}t_0^{2\alpha}t_r^{-2\alpha},
\end{equation}

\begin{equation}
\label{omega-r} 
\omega(t_r)=-1+{1\over \beta}+{\alpha \over \beta}+{2\alpha \over \beta^2}-{1\over \beta^2}{8\pi \rho_1 \over \phi_0 a_0^4}{4\over 3}t_0^{4\alpha+\beta}t_r^{2-4\alpha-\beta}+{1\over \beta^2}{2k \over a_0^2}t_0^{2\alpha}t_r^{2(1-\alpha)},
\end{equation}

\noindent where the subindex $r$ indicates the radiation time. From the assumed functional form for $a(t)$ given by Eq. (\ref{a-phi}), we have $q>0$ ({\it i.e.} decelerated expansion) for $\alpha <1$. Some exact solutions for a radiation fluid have been found in Brans-Dicke theory (see {\it e.g.} \cite{lambda1}, \cite{LP}  \cite{chauvet}), where $a(t) \sim t$ and consequently $q=0$. As we have mentioned in the introduction, one of the problems in BD theory is the difficulty for describing a decelerated expansion for early epochs, in contradiction with predictions of big-bang nucleosynthesis, unless $\omega$ be allowed to change with time (see \cite{diego}). Thus, in the present theory where $\omega$ is a function of the time, we find $a(t_r) \sim t_r^{\alpha}$, where $\alpha<1$ is required in order to have a decelerated expansion. Additionally, we expect that $\phi(t_r) \sim t_r^{\beta}$ with $\beta <0$, and $\Lambda(t_r)$ be a decreasing function, which from Eq. (\ref{lambda-r}) implies $4\alpha+\beta>0$. 

In Figure 2, we plot $\Lambda(t_r)$ as given in Eq (\ref{lambda-r}) for a flat model ($k=0$) and for some values of $\alpha$ and $\beta$ which fulfill the conditions of a decelerated expansion, a decreasing scalar field function $\phi(t_r)$, and a decreasing cosmological function $\Lambda(t_r)$. 

In figure 3, we plot $\omega(t_r)$ as given in Eq. (\ref{omega-r}) for the same values of $\alpha$ and $\beta$ used in Fig. 2. Note that the scale of the {\it y} axis in panels (b)-(d) of Fig. 3, shows that $\omega$ change so little that we can assure that $\omega$ remains practically constant in the interval from $1$ to $10^6 \, {\rm s}$. Panel (a) shows that for the values $\alpha=0.5$ and $\beta=-1/4$, $\omega$ is a slowing decreasing function (it decreases $1/10$ in $10^6$ s). Note also that for some values of $\alpha$ and $\beta$ in the interval of interest, $\omega$ get negative values. 

For Fig. 2 and Fig. 3 we have assumed $H_0=65 \, {\rm km\, \, s^{-1}\, \, Mpc^{-1}}$, and $8\pi \rho_1 /\phi_0 \sim 7.4 \times 10^{-36} \, {\rm s^{-2}}$ (the reason of this last value has been discussed in the previous subsection). Thus, for example, at $t_r \sim 10^{6} \, {\rm s}$, for $a(t) \sim t^{1/2}$ and $\phi(t) \sim t^{-1/4}$ we get $\Lambda(t_r) \sim 9 \times 10^{-14} {\rm s}^{-2}$ and $\omega \sim 9$. For  $a(t) \sim t^{0.6}$ and $\phi(t) \sim t^{-2}$ we get $\Lambda(t_r) \sim 48 \times 10^{-14} {\rm s}^{-2}$ and $\omega \sim -1.5 $. Then for $\alpha$ and $\beta$ in the required interval for a decelerated expansion, we find that $\Lambda(t_r)$ has larger values than $\Lambda(t_0)$ and $\omega(t_r)$ is smaller than $\omega(t_0)$, which is consistent with the expectations of the present theory in the sense that the effective cosmological constant $\Lambda(\phi)$ has a large value in the past and a small value today (of the order of the value reported by measurements), as well as the coupling function $\omega(\phi)$ which starts with small values at early epochs and becomes larger today (consistent with the weak field limits).

We can also estimate the values of the parameters $\Omega_\Lambda$ and $\Omega_\phi$ at the radiation epoch, using Eq.(\ref{a-phi}) in Eqs. (\ref{omega-lamb-q})-(\ref{omega-phi-q})

\begin{equation}
\label{om-lam-r}
\Omega_\Lambda|_r={1\over 3\alpha}\left[ {\beta^2\over 2}-{\beta \over 2}+3\alpha^2-\alpha +{5 \over 2}\alpha \beta \right]-{8\pi \rho_1 \over 2\phi_0}a_0^{-3} {2\over 9\alpha^2}H_0^{-\beta-4\alpha}t_r^{2-\beta-4\alpha}, 
\end{equation}

\begin{equation}
\label{om-phi-r}
\Omega_\phi|_r=-{\beta \over \alpha}+{1\over 6\alpha^2}\left[ \beta(1-\beta)+2\alpha+\alpha\beta \right]-{8\pi \rho_1 \over 2\phi_0}a_0^{-3}{4\over 9\alpha^2}H_0^{-\beta-4\alpha}t_r^{2-\beta-4\alpha}.
\end{equation}

\noindent Using again $H_0=65 \, {\rm km\, \, s^{-1}\, \, Mpc^{-1}}$, and $8\pi \rho_1 /\phi_0 \sim 7.4 \times 10^{-36} \, {\rm s^{-2}}$, we calculate the values of $\Omega_\Lambda$ and $\Omega_\phi$ for the same values of $\alpha$ and $\beta$ used in Fig. 2 and Fig. 3. The results are shown in Fig. 4 and Fig. 5. We see that $\Omega_\Lambda$ and $\Omega_\phi$ remains practically constant in the interval from $1$ to $10^6$ s. We note also that $\Omega_\Lambda + \Omega_\phi \sim 1$ in each case.


\begin{figure}[hbp]
\begin{center}

\includegraphics[width=220pts]{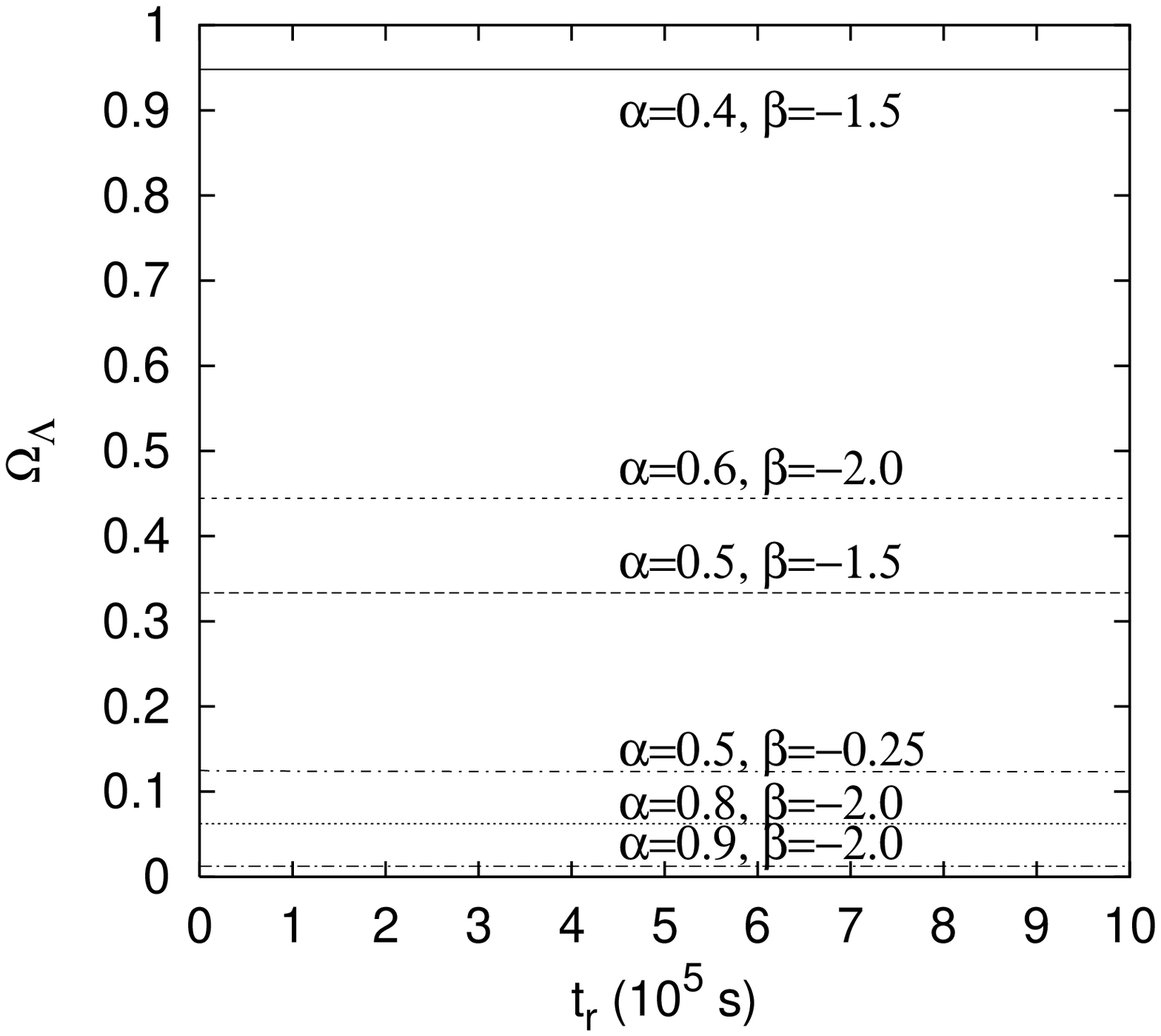}{}

\end{center}

\vspace*{8pt}

\caption[]{$\Omega_\Lambda$ {\it vs} $t_r$ from Eq. (\ref{om-lam-r}) for different values of $\alpha$ and $\beta$ which satisfy the restrictions $4\alpha+\beta>0$, $\alpha<1$, and $\beta<0$.}

\end{figure}


\begin{figure}[!htp]
\begin{center}

\includegraphics[width=220pts]{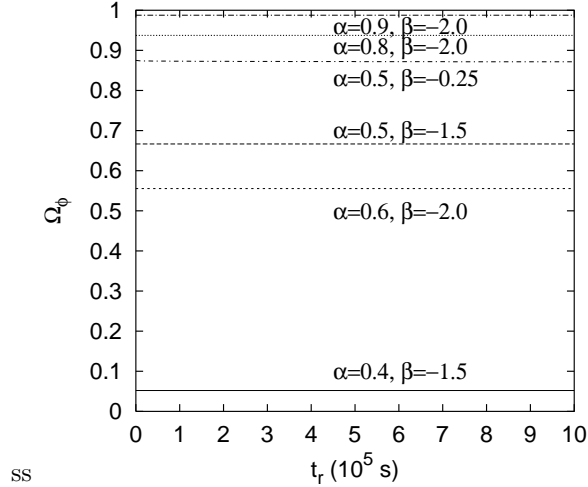}{}

\end{center}

\vspace*{8pt}

\caption[]{$\Omega_\phi$ {\it vs} $t_r$ from Eq. (\ref{om-phi-r}) for different values of $\alpha$ and $\beta$ which satisfy the restrictions $4\alpha+\beta>0$, $\alpha<1$, and $\beta<0$.}

\end{figure}

\section{Summary and final remarks}

We have studied a homogeneous and isotropic cosmological model in the context of a generalized scalar-tensor theory. The coupling function between gravity and scalar field $\omega(\phi)$, as well as the effective cosmological constant $\Lambda(\phi)$ are both functions of the scalar field (or the time) and they have been let as undetermined functions. 

Our system of field equations contains more unknowns than independent differential equations, thus instead of writing the solutions for the scale factor $a(t)$ and the scalar field $\phi(t)$ in terms of the functions $\omega(\phi)$ and $\Lambda(\phi)$, we have gotten expressions for $\omega(\phi)$ ( Eq.(\ref{omega-q})) and $\Lambda(\phi)$ (Eq.(\ref{lambda-q})), in terms of $a(t)$ and $\phi(t)$. The advantage of proceeding in this way is that we have been able to get a set of conditions of this theory that agree with physical expectations and observational results.

The obtained expression for $\omega(\phi)$ shows that, in the most general case, it is not a simple function of the scalar field, but a function of first and second temporal derivatives of the scalar field, as well as the scale factor.

We have obtained expressions for  $\Omega_m$ (Eq.(\ref{omega-m})), $\Omega_\Lambda$ ( Eq.(\ref{omega-lamb-q})), and $\Omega_\phi$ (Eq.(\ref{omega-phi-q})) which in scalar-tensor theories (STT), are equivalent parameters to the energy density contributions $\Omega_m$ and $\Omega_\Lambda$, defined in General Relativity (GR). Our expressions of $\Omega_m$ and $\Omega_\Lambda$ in STT are reduced to those well known expressions in the limit of GR. 
We have got a general expression for the deceleration parameter $q$ (Eq.(\ref{q-omegas-p})), from which it is clear that $q_0$ (Eq.(\ref{q0-omegas}))  will be negative (or positive) depending not just on the equation of state and $\Omega_{m0}$, but on the contributions of the scalar field through $\Omega_{\Lambda_0}$ and $\Omega_{\phi_0}$. 
We have shown here that it is necessary to consider some assumptions that complete the system of equations and look for conditions which satisfy the physical requirements of the weak field limits. With this purpose, we have assumed a power law function of the time (Eq.(\ref{a-phi})) for both, the scale factor $a \sim t^\alpha$  and the scalar field $\phi \sim t^\beta $. Using this assumption, we have discussed the conditions for an accelerated expansion at the present time $t=t_0$, as well as the correct limits for $\omega(\phi)$ in the weak  field limit. 

We have found that in order to satisfy the weak field limits, it is necessary to have $\beta \sim 0$ at $t=t_0$, but a different value of $\beta$ should be assumed for $t \neq t_0$ (see discussion under Eq.(\ref{a-phi})). In order to overcome this problem, we have assumed two regions of the function $\phi(t)$ given in Eq.(\ref{a-phi}), one for $\beta \sim 0$ at $t \sim t_0$ and another for $\big |\beta\big | \neq 0$ at an early epoch. 

With those conditions we also found a range of possible values for the exponent $\alpha$. We have shown that if the scale factor evolves as Eq.(\ref{a-phi}), the value of $\alpha$ (in the region of $\beta \sim 0$ and $t \sim t_0$), depends on the equation of state through $\gamma$ (see Eq.(\ref{density})). Thus, if the present universe would be ruled by a dust state equation and the acceleration of the universe is due to the scalar field, with $\Omega_{m0}\sim 0.33$, $\Omega_{\Lambda_0}+\Omega_{\phi_0} \sim 0.66 $, $\Lambda(t_0)\sim 10^{-35} \, {\rm s}^{-2}$,  $\omega(t_0) \longrightarrow \infty$ and ${\dot \omega \over \omega^3}{1\over \dot \phi}|_{t=t_0}=0$, then we found that $q_0 \sim -0.14$ or $a(t) \sim a_0 (t/t_0)^{7/6}$.

We also have studied the case of a model dominated by radiation. We got expressions at the radiation time $t_r$, for $\Lambda$ and $\omega$ in terms of $\alpha$, $\beta$, $t_0$, and constants. We do not have a restriction value on $\Lambda(t)$ for this epoch, but we expect that $\Lambda(t_r)$ has a larger value than $\Lambda(t_0)$. We are dealing with a time dependent cosmological ``constant'' such that it can have a larger value in a early universe and it will decrease with time until the small value estimated today. We also expect that $\omega(t)$ gets a small value at an epoch dominated by radiation and continues growing until the value measured today in solar system measurements, (and agrees with weak field limits). We have found a range of values for $\alpha$ and $\beta$ which fullfill all this conditions and a decelerated expansion in this epoch. We have calculated $\Lambda$, $\omega$, $\Omega_\Lambda$ and $\Omega_\phi$ for some values of $\alpha$ and $\beta$ in the interval found here. The obtained results are consistent with the expectactions of this theory.

It is clear that the previous results depend strongly on the form of the assumed functions for $a(t)$ and $\phi(t)$, specially the values of the power law constants. The age of the universe also depends on the assumed function for $a(t)$, which is given by $t_0=\alpha/H_0$, where $H_0$ is the Hubble parameter measured today. We found that an accelerated expansion requires $\alpha >1$, thus $t_0 > 15$ Gyr, which is a large value if we accept that $t_0 \sim 13.4 \pm 1.6$ Gyr \cite{line}. Because the larger the value of $\alpha$ (or smaller acceleration), the older the universe is, the assumed function for $a(t)$ could be troublesome. 

Even if there is not certainty that these are the actual functions for $a(t)$ and $\phi(t)$, the obtained results show that it is possible to get a set of conditions in STT, in order to get an accelerated expansion at the present epoch, and a decelerated expansion in a remote past, and also have a good agreement with values of cosmological parameters. The present results show the kind of information which we require and the possible results which we can obtain from the study of these problems in generalized scalar-tensor theories.

Some important aspects still need to be discussed in the context of the theory which we have presented here, for example, the study of density perturbations, in order to know if the conditions for structure formation in generalized scalar-tensor theories are consistent with the results obtained in this work. This aspect will be discussed in a future work.

\section*{Acknowledgments}

One of us (LMDR) wants to thank Remigio C. Trujillo for useful comments, to CONACYT for financial support and the Physics Department at the University of Florida for its hospitality.


\end{document}